\documentclass[prl,twocolumn,amsmath]{revtex4-1}

\usepackage{graphicx, color}
\usepackage{amssymb,amsfonts,amsmath}

\def\s{\sigma}

\def \g{\gamma}    
\def \w{\omega}     
\def \s{\sigma}      
\def \e{\epsilon}    
   \def \d{\delta} 
    \def \l{\lambda}

\def \mat{\text}

\def \del{\partial}    
 
\def \hf{\tfrac{1}{2}} 
\def \HF{\dfrac{1}{2}}  

\def \>{\rangle} 
\def \<{\langle} 
 
\def\be{\begin{equation}} 
\def\ee{\end{equation}} 
\def\longrightharpoonup{\relbar\joinrel\rightharpoonup}
\def\longleftharpoondown{\leftharpoondown\joinrel\relbar}

\def\longrightleftharpoons{
  \mathop{
    \vcenter{
      \hbox{
      \ooalign{
        \raise1pt\hbox{$\longrightharpoonup\joinrel$}\crcr
	  \lower1pt\hbox{$\longleftharpoondown\joinrel$}
	  }
      }
    }
  }
}

\newcommand \bea {\begin{eqnarray}} 
\newcommand \eea {\end{eqnarray}}

\begin{document}

\title{A phase transition between the niche and neutral regimes in ecology}

\author{Charles K. Fisher}
\affiliation{Department of Physics, Boston University, Boston, MA 02215}

\author{Pankaj Mehta}
\affiliation{Department of Physics, Boston University, Boston, MA 02215}

\begin{abstract} 

An ongoing debate in ecology concerns the impacts of ecological drift and selection on community assembly. Here, we show that there is a sharp phase transition in diverse ecological communities between a selection dominated regime (the niche phase) and a drift dominated regime (the neutral phase). Simulations and analytic arguments show that the niche phase is favored in communities with large population sizes and relatively constant environments, whereas the neutral phase is favored in communities with small population sizes and fluctuating environments. Our results demonstrate how apparently neutral populations may arise even in communities inhabited by species with varying traits. 

 \end{abstract}

\maketitle

\section{Significance }
In recent years, there has been a vigorous debate among ecologists over the merits of two contrasting models of biodiversity: the “niche” and “neutral” theories of ecology. Using two different theoretical models of ecological dynamics, we show that there is a sharp phase transition between a selection dominated regime (the niche phase) and a drift dominated regime (the neutral phase). This is analogous to the phase diagram of water, which can be in the solid, liquid, or gas phases depending on the temperature and pressure. Our results demonstrate how the niche and neutral theories both emerge from the same underlying ecological principles.

\section{Introduction}
The success of the neutral theory of biodiversity and biogeography  \cite{rosindell_case_2012, hubbell_unified_2001}  at explaining patterns in biodiversity has resulted in a vigorous debate on the processes underlying the assembly, dynamics, and structure of ecological communities \cite{rosindell_case_2012,ricklefs_global_2012,wootton_field_2005,mcgill_test_2003,ricklefs_unified_2006,dornelas_coral_2006,jeraldo2012quantification,tilman_niche_2004,volkov2009inferring,haegeman2011mathematical,chisholm_niche_2010}. Starting with the pioneering work of MacArthur \cite{macarthur_limiting_1967, chase_ecological_2003, tilman_resource_1982}, ecologists have emphasized the roles of interspecific competition and environmental interactions in community assembly and dynamics. These niche-based models emphasize ecological selection as the driving force of community assembly, whereas neutral models of biodiversity assume a functional equivalence between species and emphasize the role of ecological drift (i.e.\ stochasticity) in community dynamics  \cite{rosindell_case_2012, hubbell_unified_2001, volkov_neutral_2003, azaele_dynamical_2006}. The success of both types of models at explaining ecological data highlights the crucial need for understanding the impacts of ecological drift and selection in community ecology.

\subsection{Hypothesis}

\begin{figure*}[t]
\includegraphics[width=6.5in]{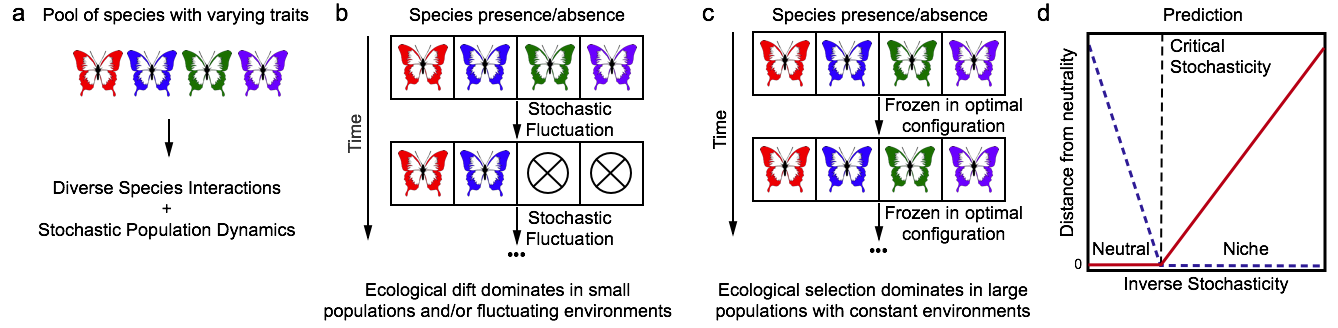}
\caption{ A schematic illustrating the intuition underlying our hypothesis for a phase transition between the neutral and niche regimes in ecology. a) The important ingredients of our model are a large pool of diverse species, implying a diversity of species interactions, subject to stochastic population dynamics. b) Stochastic ecological drift will dominate the dynamics of communities with small population sizes and/or fluctuating environments. c) By contrast, stabilizing selective forces will cause a community with a large population size and a constant environment to ``freeze'' into a unique, optimal configuration. d) We predict that the transition between the drift dominated (neutral) phase and the selection dominated (niche) phase is sharp. That is, the community behaves exactly neutral when the inverse stochasticity is less than a critical threshold, and the deviation from neutrality rises quickly once the inverse stochasticity is larger than the critical threshold. The red line represents an order parameter based on the distance from neutrality, the dashed blue line represents an order parameter based on the niche phase, and the dashed black line denotes the critical stochasticity.
\label{fig:fig1}}
\end{figure*}

We begin with a hypothesis that a diverse ecological community with many species  can be either neutral or non-neutral depending on the state of its environment. We call the regime in which a community is well described using neutral models the ``neutral phase'', and the regime in which the community behaviors are inconsistent with neutrality,  the ``niche phase''. The dynamics in the neutral phase are dominated by stochasticity whereas the dynamics in the niche phase are dominated by selection.  Our goal in this paper is to demonstrate that these two phases naturally emerge from simple probabilistic models of ecological dynamics, and that a community may abruptly transition from one phase to the other as its environment is altered (see Fig.\ \ref{fig:fig1}).  

Historically, ecological neutrality is based on the assumption of functional equivalence, which states that trophically similar species are essentially identical in terms of their vital characteristics, such as birth and death rates \cite{hubbell2005neutral}. Ecological neutrality, however, is generally not a measurable feature of a community. Therefore, we will adopt a pragmatic definition of neutrality: we say that a community is ``statistically neutral'' if its multivariate distribution of species abundances cannot be distinguished from a distribution constructed under the assumption of ecological neutrality.  In other words, the species abundance distributions of statistically neutral communities are indistinguishable from those of communities of functionally identical species. Note that ecological neutrality implies statistical neutrality, but statistical neutrality does not necessarily imply ecological neutrality. We can now restate our hypothesis more precisely: as the characteristics of an ecosystem change (e.g. carrying capacity, immigration rate), there will be a sharp transition between a neutral phase where the ecosystem behaves as if it is effectively neutral and a niche phase where the multivariate species distribution is inconsistent with statistical neutrality.

\subsection{Background on phase transitions in disordered systems}

Our hypothesis that ecological systems are likely to exhibit multiple phases is based on an analogy with disordered systems in physics. For this reason, we briefly provide some background on phase transitions in disordered systems. A phase transition refers to an abrupt change in the qualitative behavior of a system as one of its characteristics, or a characteristic of its environment, is altered \cite{kadanoff_statistical_2000}. The most well-known example may be the behavior of water, which can be found as a solid, liquid, or gas depending on temperature and pressure. Disordered systems often display a more complicated type of phase transition, labeled the freezing transition, where the system configuration gets ``frozen'' into a particular state. 

One illustrative example of a disordered physical system is a protein \cite{bryngelson1987spin}.   A protein can be thought of as a disordered system in which the different amino acids along the protein chain interact heterogeneously. The diversity of interactions in a protein distinguish natural proteins from homopolymers, and is what allows some proteins to fold to a stable native structure, while causing others (like prions or amyloids) to misfold. In ecology, this is analogous to the observation that the diversity of interactions between the species in a community distinguish niche-like communities from neutral communities. To continue our analogy, at high temperatures, a typical polypeptide sequence will be in an unfolded phase where it samples different configurations randomly. If the temperature is lowered below a critical value, the polypeptide will freeze into a single structure (the folded state). This phase transition occurs when the stochasticity impacting the dynamics (i.e.\ the temperature) is smaller than the energetic differences in the interactions of the amino acids along the chain. If we take this analogy seriously,  we should expect to find a critical amount of stochasticity, compared to the diversity of species traits, that separates neutral and niche communities.

\section{Theoretical models for studying community assembly}
To test our hypothesis regarding the niche-to-neutral phase transition, we analyzed two models of ecological dynamics: (1)  a generalized Lotka-Volterra (LV) system including immigration and stochasticity and (2)  a new binary model  for the the presence/absence (PA) of the species in a community. Each of these models has advantages and disadvantages.  The LV model is a widely-used and interpretable model of many ecological phenomena. However, in general, it is intractable to perform analytic calculations using the LV model and one must rely on numerical simulations. In contrast to the LV system, the PA model is amenable to analytical arguments but this comes at the expense of ignoring species abundances. Both models assume well-mixed populations, though relaxing this assumption is an important avenue of future research. These two models correspond to extreme cases of functional responses \cite{holling1959some,abrams1987functional}. The functional response in the PA model is essentially a step function in which species only interact when their abundances are above a threshold. By contrast, the LV model corresponds to linear functional responses. We expect that real communities lie somewhere in between these models. 

\subsection{Parameterizing ecosystem characteristics}

In order to construct ecological phase diagrams, it is necessary to parameterize ecosystem characteristics. Since we are interested in stochastic community assembly, we must introduce parameters that reflect the impact of stochasticity as well as parameters that capture variation in species traits. Due to the similarity of the two models, we will use the same symbols for analogous parameters with an added tilde for parameters in the PA model (e.g.\ $K$ denotes carrying capacity in the LV model and $\tilde{K}$ denotes carrying capacity in the PA model). 

There are two potential sources of stochasticity in the ecological dynamics: ``demographic stochasticity''  resulting form  random births and deaths in small populations and  ``environmental stochasticity'' caused by random variations in the environmental conditions.  While there is no doubt that the origin of the stochasticity is important for making quantitative ecological predictions \cite{lande2003stochastic},  extensive numerical simulations suggest that the qualitative phase diagrams are insensitive to these details (Supporting Information). For this reason, we parameterize the amount of stochasticity by a single parameter, the noise strength  $\w$ ($\tilde{\w}$). 

We must also introduce parameters describing species traits. In principle, each species in the community has a unique immigration rate, a unique carrying capacity, and some set of parameters that describes how it interacts with other species. In the main text, we will restrict ourselves to the case where all species have the same immigration rate, $\l$ ($\tilde{\l}$), and the same carrying capacity $K$ ($\tilde{K}$) (see Supporting Information for relaxation of these assumptions).  Following  May's seminal work  \cite{may_will_1972},  we  randomly draw symmetric interaction coefficients from a probability distribution and focus on describing the average behavior of ecosystems. Specifically, the interaction matrix $C$ ($\tilde{C}$) -- with element $c_{ij}$ ($\tilde{c}_{ij}$)  characterizing the strength of interaction between species $i$ and $j$ -- is drawn from a  Gamma distribution with mean $\mu/S$ ($\tilde{\mu}$/S) and variance , or ``interaction diversity'', $\s^2/S$ ($\tilde{\s}^2/S$), where $S$ is the number of species (see Supporting Information for results with other distributions). 

\subsection{Stochastic Lotka-Volterra dynamics}
The first model that we will analyze is a system of stochastic Lotka-Volterra (LV) equations including immigration. Niche-based models of community assembly frequently employ LV equations as a simplified description of ecological dynamics within a well-mixed community \cite{macarthur_limiting_1967,macarthur1970species,chesson_macarthurs_1990,chesson2008interaction}.  Here, we study a system of LV equations incorporating immigration and multiplicative noise (i.e.\ stochasticity). The rate of change in the abundance ($x_i$) of species $i=1,\ldots,S$ is:
\be
\label{eq:LV}
\frac{d x_i}{dt} = \l + x_i(K - x_i) - \sum_{j \neq i} c_{ij} x_i x_j + \sqrt{\w x_i} \eta_i(t)
\ee
The first term ($\l$) is the rate that species $i$ immigrates into the local community from an infinitely large regional species pool. The second term ($x_i(K - x_i)$) limits the population of species $i$ to its carrying capacity ($K$) in the absence of immigration and species interactions. The third term ($\sum_{j \neq i} c_{ij} x_i x_j $) describes the effects that other species in the community have on species $i$ according to their interaction coefficients ($c_{ij}$). All of these deterministic terms (i.e.\ $\l$, $K$, and $c_{ij}$) collectively represent the effects of ecological selection on the abundance of species $i$. Ecological drift is incorporated into our model through the last term ($\sqrt{\w x_i} \eta_i(t)$), which represents stochasticity using a Gaussian ``white noise'' $\eta_i(t)$, with mean $\<\eta_i(t)\> = 0$, variance  $\< \eta_i(t) \eta_j(t') \> = \d_{ij} \d(t-t')$, and strength $\sqrt{x_i \w}$.

\subsection{Dynamics of presence/absence model}
In the PA model, a species $i$ is described by a binary variable $s_i$ with $s_i=1$ if species $i$ is  present in a community and $s_i=0$ if it is absent. The stochastic dynamics of species PA are defined by two rates: the rate at which a species immigrates into a community (i.e.\ the rate that $s_i = 0$ becomes $s_i = 1$), and the rate at which a species becomes extinct once it is in the community (i.e.\ the rate that $s_i = 1$ becomes $s_i = 0$). Thus, a species immigrates into a community and lives there for some time before it dies out, only to re-immigrate back into the community later, and so on. We assume that the rate of immigration is simply $\tilde{\l}$, and we model the rate of extinction as $\exp(- \tilde{K}(1 - \sum_j \tilde{c}_{ij}s_j) / \tilde{\w} )$. Therefore, in the absence of any interactions a species goes extinct at a rate that is exponentially slow in its carrying capacity $\exp(-\tilde{K} / \tilde{\w}  )$, and competitive species interactions effectively decrease carrying capacity through $\tilde{K}( 1- \sum_j \tilde{c}_{ij} s_j)$ \cite{macarthur_limiting_1967}. The master equation describing the dynamics of $\vec{s}$ with these rates is discussed in detail in the Supporting Information. After an initial transient period, the community reaches a steady-state where the immigration and extinction processes are balanced. Due to the simplicity of this model, we can derive an analytic expression for the steady-state probability distribution:
\be
P_{PA}(\vec{s}) = \frac{\exp( \sum_i ( \tilde{K} / \tilde{\w} + \ln \tilde{\lambda}) s_i - \frac{\tilde{K}}{2\tilde{\w}} \sum_{(i,j)} \tilde{c}_{ij} s_i s_j)}{Z(\tilde{\l}, \tilde{K}, \tilde{C}, \tilde{\w})} 
\ee
Here, $Z(\tilde{\l}, \tilde{K}, \tilde{C}, \tilde{\w})$ is a normalizing constant such that the total probability sums to one. 

\section{Measuring the neutrality of a community}
To test our hypothesis that communities can exhibit a sharp niche-to-neutral phase transition, it is necessary to define ``order parameters'' that distinguish the niche and neutral phases.  By convention, an order parameter is chosen so that it is zero in one phase, and greater than zero in the other. Recall that the dynamics in the neutral phase are dominated by stochasticity and species abundance distributions in this phase are indistinguishable from those obtained from a neutral model with functionally equivalent species. By contrast, the niche phase is dominated by interactions and species abundance distributions are peaked around the equilibrium value they would have in the absence of stochasticity.

Using these intuitions we can define order parameters for both the LV model and the PA model. In the LV model, we define an order parameter that measures the distance (i.e.\ Kullback-Leibler divergence) between the multivariate species distribution resulting from LV dynamics and the multivariate species distribution resulting from purely neutral dynamics (see below). This order parameter is zero in the neutral phase and non-zero in the niche phase. For the PA model, it is convenient to consider a different order parameter, the Shannon entropy, of the steady-state PA probability distribution. The Shannon entropy is zero in the niche phase and non-zero in the neutral phase. We now discuss both of these order parameters in more detail.

\subsection{Measuring neutrality in LV Models}
Early studies attempting to quantify the neutrality of a community focused on the shape of the marginal species abundance distribution, i.e.\ a histogram indicating the number of species with 10 individuals in the community, the number of species with 20 individuals in the community, and so on. However, recent studies have shown that both and neutral and non-neutral ecological models give rise to similar marginal species distributions \cite{chisholm_niche_2010}. For this reason, to measure neutrality in the LV model we will  utilize the multivariate distribution of species abundances.  In particular, we quantify statistical neutrality in our LV simulations by measuring the distance between the steady-state distributions of species abundances obtained from the LV model ($P_{LV}(\vec{x})$) and purely neutral dynamics  ($P_N(\vec{x})$). The measure of distance that we will use is called the Kullback-Leibler divergence $\text{KL}(P_{LV} || P_{N} ) = \int d \vec{x} P_{LV}(\vec{x}) \ln P_{LV}(\vec{x}) / P_N(\vec{x})$ \cite{kullback_information_1951}. One interpretation of $\text{KL}(P_{LV} || P_{N} )$ is as the amount of information about the true multivariate species abundance distribution (i.e.\ $P_{LV}(\vec{s})$) that is lost by approximating the distribution with one obtained from a neutral model (i.e.\ $P_{N}(\vec{x})$). The KL-divergence ranges from zero to infinity, with $\text{KL}(P_{LV} || P_{N} ) = 0$ implying that the simulated distribution is identical to the distribution obtained under the assumption of neutrality. We study the average of the KL-divergence over many random realizations of the species interactions, i.e.\ $\< \text{KL}(P_{LV} || P_{N} ) \>$. We expect  $\< \text{KL}(P_{LV} || P_{N} ) \>  \approx 0$ in the neutral phase, whereas $\< \text{KL}(P_{LV} || P_{N} ) \> \gg 0$ in the niche phase. Similar results are obtained with distance measures other than the KL-divergence (Supporting Information).

\begin{figure*}[t]
\includegraphics[width=6.5in]{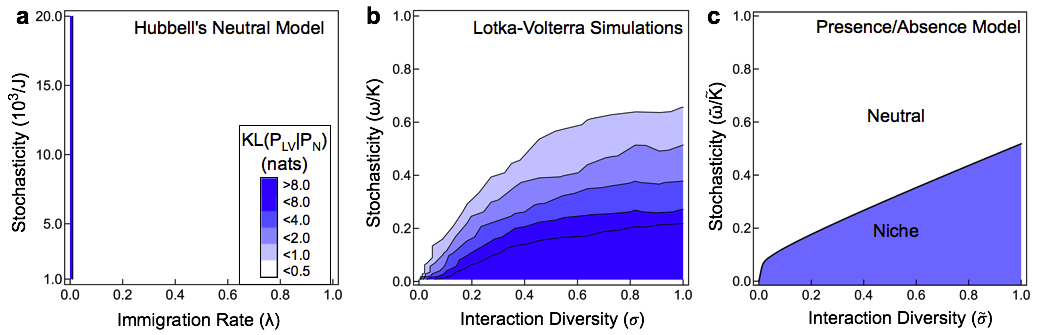}
\caption{ Phase diagram of neutral and competitive ecosystems. a) Communities simulated according to Hubbell's neutral model are statistically neutral a KL-divergence equal to zero for all positive immigration rates and community sizes (J). b) Simulations of competitive LV communities with immigration display two phases: a statistically neutral phase with $\<\text{KL}(P_{LV} || P_N) \> \approx 0$  and a niche phase with $\<\text{KL}(P_{LV} || P_N) \> \gg 0$. Note that the colors represent exponential growth in the KL-divergence. The critical stochasticity defining the phase boundary scales with interaction diversity ($\s$). Simulations were performed with $\mu = 1.0$ and $\l = 0.01$. c) The phase diagram calculated from the presence/absence model has a statistically neutral phase and a niche phase, and a phase boundary that scales with interation diversity ($\tilde{\s}$). The phase diagram was calculated with $\tilde{\mu} = 1.0$ and $\tilde{\w} \ln \tilde{\l} = \tilde{K}/2$.  
\label{fig:fig2} }
\end{figure*}

In principle, it is possible to use an explicit formula for $P_N(\vec{x})$ from a specific neutral ecological model. However, many variations of neutral ecological models have been proposed and it is unclear which neutral model to use to calculate our order parameter. To circumvent this problem, we exploit the observation that the multivariate species abundance distributions of all neutral models share several features. Since we have restricted ourselves to considering LV systems where all species have the same immigration rate, we will also restrict ourselves to considering neutral models where this assumption holds. The implications of non-uniform immigration rates are discussed in the Supporting Information. With this caveat in mind, we observe that ecologically neutral models are also statistical neutral. Namely, the time-averaged moments of the abundance of species $i$ are the same as the time-averaged moments of the abundance of species $j$. Moreover, the correlation in the abundances of species $i$ and $j$ is the same as the correlation in the abundances of species $k$ and $l$ (see Supporting Information). Simulations shown in Fig.\ \ref{fig:fig2}a demonstrate that this is the case, at least for Hubbell's neutral model (Materials and Methods) where the KL-divergence equals zero for all positive immigration rates.  Finally, we note that although ecological neutrality implies statistical neutrality, statistical neutrality does not necessarily imply ecological neutrality. Thus, our use of statistical neutrality is consistent with the interpretation of ecological neutrality as a type of null model that allows one to identify communities in which selection is important. 

\subsection{Measuring neutrality in the presence/absence model}
In the PA model, we do not have access to species abundances. For this reason, it is convenient to define a different order parameter that measures the fluctuations in the binary vector of community composition, $\vec{s}$:  the entropy,  $H[P_{PA}] = -\sum_{\vec{s}} P_{PA}(\vec{s}) \ln P_{PA}(\vec{s})$, of the steady-state probability distribution $P_{PA}(\vec{s})$. In the absence of stochasticity, $\vec{s}$  will "freeze" into a unique configuration resulting from ecological selection and  $H[P_{PA}] = 0$. In contrast, if the dynamics are entirely random then each species will randomly flip between being absent and present in the community and  $H[P_{PA}] = S \ln 2$. For diverse ecosystems with $S \gg 1$, we can define the boundary between the neutral phase and the niche phase as the points where $\< H[P_{PA}] / S \> = 0$, with angular brackets denoting  averaging over random realizations of interaction coefficients.

\section{Phase diagrams for ecological dynamics}

\subsection{Phase Diagrams}

Armed with the order parameters discussed in the last section, we can construct phase diagrams for both the LV and PA models. Fig.\ \ref{fig:fig2} shows the KL-divergence and entropy as a function of stochasticity and interaction diversity for the two models. First, we note that the phase diagram determined using LV simulations is remarkably similar to the phase diagram calculated using our PA model (compare Figs.\ \ref{fig:fig2}b and c), which suggests that our results are fairly robust to model details. Figure \ref{fig:fig2}b shows that there is a large neutral regime in which $\< \text{KL}(P_{LV} || P_{N} ) \>  \approx 0$ in the LV simulations. The distance from neutrality rises quickly once the stochasticity is lowered below a critical value. That is, $\< \text{KL}(P_{LV} || P_{N} ) \> $ increases rapidly for small $\w / K$; note that the colors in Fig.\ \ref{fig:fig2}b represent exponential growth in $\< \text{KL}(P_{LV} || P_{N} ) \> $.

Figure \ref{fig:fig2}c shows the phase diagram for the PA model. In the limit the number of species S becomes large, the entropy is strictly zero in the niche phase (blue shaded area) and different from zero in the neutral phase (white area).  In particular, we find that the PA of the species in a community freezes into a small number of configurations determined by the species traits if the stochasticity ($\tilde{\w}$) is lowered below a critical value. This freezing is indicative of a phase transition from neutrality to niche dominated ecological dynamics.

\begin{figure}[b]
\begin{center}
\includegraphics[width=3in]{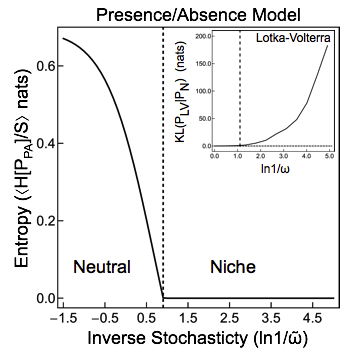}
\caption{ Sharpness of the transition between the niche and neutral phases. Note that the order parameters for the two models are different: the order parameter for the PA model is zero in the niche phase and greater than zero in neutral phase, whereas the order parameter for the LV model is greater than zero in the niche phase and zero in the neutral phase. The average entropy $\<H[P_{PA}]/S\>$, which is a measure of fluctuations in the community composition, is positive in the neutral phase and zero in the niche phase illustrating the freezing transition in the PA model. (Inset) In LV simulations the distance from neutrality $\<\text{KL}(P_{LV} || P_N) \>$ is esstentially zero in the neutral phase, and quickly rises to large values in the niche phase. Parameters: $\tilde{\mu} = 1.0$, $\tilde{\sigma} \approx 0.4$, $\tilde{\w} \ln \tilde{\lambda} = 3 \tilde{K} /2$, $\mu = 1.0$, $\sigma \approx 0.4$, $\lambda = 0.01$. 
\label{fig:fig3} }
\end{center}
\end{figure}

\subsection{Scaling relation for the niche-neutral phase boundary}

We can explicitly calculate the phase boundary separating the niche and neutral phases using the PA model. For diverse ecosystems with many species $S \gg 1$,  the relation defining the phase boundary can be derived by mapping the problem to the Random Energy Model in physics \cite{derrida_random-energy_1980,derrida_random-energy_1981} (see Supporting Information). Using this mapping we can derive a simple scaling relation that indicates when an ecological community will transition between the niche to neutral phases (see Supporting Information and Fig.\ S2):
\be
\frac{ \text{stochasticity} }{ \text{carrying capacity}} \sim \frac{ \text{immigration}  \times \text{interaction diversity} } {\text{mean interaction strength} } \nonumber
\ee
The niche phase is favored when the interaction diversity is large relative to the impact of stochasticity on the dynamics of the population. By contrast, the neutral phase is favored when the interaction diversity is small relative to the impact of stochasticity on the dynamics of the population. This confirms the basic intuitions about ecological dynamics that were suggested by the analogy with protein folding discussed in the introduction.

\begin{figure*}
\includegraphics[width=6.5in]{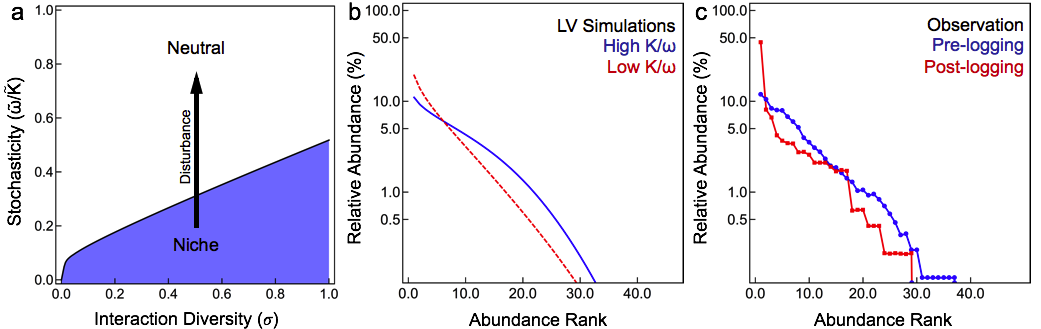}
\caption{ Temporal variation in stochasticity and biodiversity in disturbed habitats. a) An environmental disturbance that decreases carrying capacity may cause a community to shift from the niche phase to the neutral phase. b) A community with a high carrying capacity ($K = 1.0$: blue) has a less skewed species abundance distribution than a community with a low carrying capacity ($K = 0.1$: red), as shown by the steeper red curve in the rank-abundance plot obtained from LV simulations. Simulations were run with $\mu = 1.0$, $\s = 0.5$, $\l = 0.02$, and $\w = 0.6$. c) Similarly, rank-abundance plots of butterfly species in a tropical Indonesian forest before (blue) and after (red) logging reflect an increase in skewness of the species abundance distribution following the disturbance \cite{hill_effects_1995}.  
\label{fig:fig4} }
\end{figure*}

\subsection{On the sharpness of the transitition}

The transition between the niche and neutral phases is fairly sharp (see Fig.\ \ref{fig:fig3}). In the LV model the distance from neutrality ($\< \text{KL}(P_{LV} || P_{N} ) \>$) increases rapidly when the stochasticity is lowered below the critical value.  In the PA model, the derivative of the entropy with respect to stochasticity is undefined along the phase boundary, the signature of a freezing phase transition in the theory of disordered systems. Comparing the two models, the niche-to-neutral transition in the PA model appears to be sharper than in the LV model.  This difference arises due to the differences in the functional responses of the two models. These two models were chosen, in part, because they  represent the  two extremes of possible species functional responses (linear versus step-function). We expect the functional responses of real ecological communities lie somewhere in between these two models. For this reason, we expect that real ecological communities will also exhibit a sharp transition between the niche and neutral phases.

\section{Ecological Implications}

\subsection{The prevalence  of neutral and niche communities}

Our model suggests that neutral communities and niche-like communities both correspond to large volumes of the ecological phase diagram. Moreover, our model is such that species always have real differences in traits, but these differences in species traits leave no trace on the equilibrium distribution of species abundances in the statistically neutral regime. This does not preclude the possibility that one could observe the effects of species trait variation on other types of observations. This may explain the success of neutral models at explaining many large-scale patterns in ecology, even though selective forces are well-documented, and ubiquitous, on local scales. Furthermore, because the transition between the niche and neutral phases is sharp, the crossover region surrounding the phase boundary corresponding to ``nearly neutral'' communities occupies only a small volume of the phase diagram. As a result, we predict that nearly neutral communities should actually be quite rare, so long as there is not an external force (e.g.\ group selection) driving communities towards the niche-neutral boundary. 

\subsection{Ecological disturbances}

One of our main predictions is that the apparent neutrality of an ecological community is a function of both the inhabiting species and the environment. As a result, it is possible to drive a community between the niche and neutral phases by changing the environmental conditions. As an example,  we consider the effects of selective logging on a population of butterflies in a tropical forest on Buru, Indonesia \cite{hill_effects_1995}. Through habitat destruction, logging essentially moves the butterfly community from a position with high $K/\w$ to one with low $K/\w$, tracing a path along the stochasticity axis in the phase diagram (Fig.\ \ref{fig:fig4}a). Our model predicts that when a diverse community within the niche phase is placed under a stress that lowers $K/\w$ to the critical value, it will undergo a phase transition to the neutral regime. LV simulations show that this transition results in a collapse of biodiversity and leads to an increase in the skewness of the species abundance distribution (Fig.\ S4). The increase in skewness of the species abundance distribution calculated from LV simulations is evident in a steeper curve in the rank-abundance plot for low $K/\w$ as compared to high $K/\w$ (Fig.\ \ref{fig:fig4}b). Similarly, the observed data display an increase in the skewness of rank-abundance curve of the logged forest relative to the unlogged forest, consistent with a loss of biodiversity accompanying a niche-to-neutral transition (see Fig. \ref{fig:fig4}c). This example demonstrates the potential of ecological phase diagrams for predicting the qualitative effects of community-wide disturbances, and for capturing the characteristics that contribute to community resilience. 

\section{Conclusion}

In summary, we have argued that the niche and neutral perspectives of ecology naturally emerge from stochastic models for the dynamics of diverse populations as distinct phases of an ecological community. Population dynamics in the niche phase are dominated by ecological selection, whereas population dynamics in the neutral phase are dominated by ecological drift. Furthermore, we have derived a simple scaling relation for determining whether an ecological community will be well described by neutral models.

Our hypothesis can be experimentally tested using synthetic microbial communities in which the immigration rates, carrying capacities, and interaction coefficients can be controlled to search for a sharp transition as one moves from one region of the phase diagram to another \cite{hekstra2012contingency}. Alternatively, connections to island biogeography discussed in the Supporting Information suggest that our hypothesis could be tested by calculating the KL-divergence from the multivariate species abundance distributions on a chain of islands as a function of their distance to the mainland \cite{macarthur1963equilibrium}.

In this work, we made some simplifications that are unrealistic for natural ecological communities. For example, we restricted our analysis to well-mixed communities with purely competitive interactions. It will be necessary to generalize our results to include the effects of dispersal, mutualism, predatory-prey interactions, etc.\ in order to obtain a more quantitative model of natural communities. Nevertheless, we conjecture that the presence of a niche-neutral phase transition is robust to these model perturbations.  However, disordered systems with complex interactions display additional phases \cite{kirkpatrick1978infinite}, which suggests that more complex ecological communties may also exhibit addtional phases with novel characteristics.

\section{Numerical Simulations}
We simulated Hubbell's neutral model with a local community of $J$ individuals connected to an infinitely large metacommunity containing $S = 50$ equally abundant species. In each timestep, with probability $\l$, an individual randomly drawn from the metacommunity replaced a randomly chosen individual in the local community, or with probability $1-\l$, one randomly chosen individual in the local community replaced another randomly chosen individual in the local community. The simulations were run for $5  \times 10^{7}$ steps. Ten simulations were run for each set of parameters, and the results were averaged.

LV simulations with $S=50$ species were performed over the parameter ranges specified in the figure legends. In each case, the competition coefficients were sampled randomly, then the stochastic Lotka-Volterra equations (Eq. \ref{eq:LV}) were forward integrated for $5 \times 10^7$ steps of size $\delta t = 0.005$ using the Milstein method. Ten simulations were run for each set of parameters, and the results were averaged.

\begin{acknowledgments}
We would like to thank Jeff Gore, Alex Lang, Javad Noorbakhsh, Daniel Segre, and Les Kauffman for useful discussion and Alfred Sloan Foundation for funding.
\end{acknowledgments}

\onecolumngrid

\section{Appendix}
\setcounter{equation}{0}
\setcounter{figure}{0}
\makeatletter 
\renewcommand{\theequation}{S\@arabic\c@equation} 
\makeatletter 
\renewcommand{\thefigure}{S\@arabic\c@figure} 
\makeatletter 
\renewcommand{\theequation}{S\@arabic\c@equation} 

\section{Presence/Absence Model}

\subsection{Dynamics: Master Equation}

In this section, we introduce a phenomological model describing the probability of observing various combinations of species in a local community, which we assume is attached to a large regional species pool. The presence (or absence) of species $i \in \{1, \ldots, S\}$ is described by a binary variable $s_i$, with $s_i = 1$ if the species is present and $s_i = 0$ if it is absent. The probability of observing a particular set of present/absent species $\vec{s}$ at time $t$ is described by a probability distribution $P_t(\vec{s})$. The probability distribution is described by a differential equation called a master equation. 

The master equation describing the time evolution of  $P_t(\vec{s})$ requires us to specify two types of rates: the rate of immigration at which $s_i = 0 \rightarrow s_i = 1$, and the rate of extinction at which $s_i = 1 \rightarrow s_i = 0$. The rate of immigration of species $i$, $R_i^{I}(\vec{s}) $,  is given by:
\be
R_i^{I}(\vec{s}) = \tilde{\lambda}_i \nonumber
\ee
The rate of extinction of species $i$, $R_i^{E}(\vec{s}) $, is given by:
\be
R_i^{E}(\vec{s}) = \exp \left( -\frac{1}{\tilde{\w}} ( \tilde{K}_i - \sum_j \tilde{K}_j \tilde{c}_{ij} s_j ) \right) \nonumber
\ee
Here, $\tilde{\lambda}_i$ is the rate of immigration of species $i$, $\tilde{K}_i$ is the carrying capacity of species $i$, $\tilde{c}_{ij}$ is an interaction coefficient that describes the influence that species $j$ has on species $i$, and $\w$ describes the impact of stochasticity on species extinction. Throughout this document we are using a convention in which $\tilde{c}_{ii} = 0$ for the presence/absence model, and $c_{ii} = 0$ for Lotka-Volterra (LV) models.  The units of time have been set to the rate of extinction in the limit that $\w \rightarrow \infty$.

The time evolution of the probability distribution is described by a master equation:
\begin{align}
\label{eq:master}
\frac{d P_t (\vec{s})}{dt} & = \sum_i \{ (R_i^E(\vec{s} +  \vec{e_i} ) P_t(\vec{s} +  \vec{e_i} )  -  R_i^I(\vec{s}) P_t(\vec{s})  )( 1-s_i)  \nonumber \\
& + (R_i^I(\vec{s} -  \vec{e_i} ) P_t(\vec{s} -  \vec{e_i} )  -  R_i^E(\vec{s}) P_t(\vec{s})  ) s_i \}
\end{align}
Here, $\vec{e_i}$ is a vector with element $i=1$ and all other elements equal to zero.

The first line of Eq.\ \ref{eq:master} describes the rate of change in the probability of a community with $s_i = 0$. The positive term, $R_i^E(\vec{s} +  \vec{e_i} ) P_t(\vec{s} +  \vec{e_i} )$, reflects the process by which the probability of this state increases due to extinction events where  $s_i =  1 \rightarrow s_i = 0$. The negative term, $R_i^I(\vec{s}) P_t(\vec{s})$, reflects the process by which the probability of this state decreases due to immigration events where $s_i = 0 \rightarrow s_i = 1$. 

The second line of Eq.\ \ref{eq:master} describes the rate of change in the probability of a community with $s_i = 1$. The positive term, $R_i^I(\vec{s} -  \vec{e_i} ) P_t(\vec{s} -  \vec{e_i} )$, reflects the process by which the probability of this state increases due to immigration events where  $s_i = 0 \rightarrow s_i = 1$. The negative term, $R_i^E(\vec{s}) P_t(\vec{s}) $, reflects the process by which the probability of this state decreases due to extinction events where $s_i = 1 \rightarrow s_i = 0$. 

There is no general method for solving this equation, though it can be simulated using Gillespie's algorithm \cite{gillespie1977exact}. Below, we will show that an equilibrium solution for $P_t(\vec{s})$ can be obtained in the limit that $t \rightarrow \infty$. 

\subsection{Connection to MacArthur-Levins}

Before describing the solution for the equilibrium distribution of Eq.\ \ref{eq:master} we would first like to explain the intuition for the rates of our phenomological presence/absence model. First, it is helpful to recount the famous results on species invasion in Lotka-Volterra communities due to MacArthur and Levins \cite{macarthur_limiting_1967}. 

Deterministic models of ecological dynamics typically take the form $ d x_i / dt = \lambda_i + x_i f_i(\vec{x})$. Here, $\lambda_i$ is the rate of immigration and $f_i(\vec{x})$ is the ecological fitness of species $i$, which is a function of the population $\vec{x}$. The LV equations describe communites with linear ecological fitness $f_i(\vec{x}) = K_i - x_i -\sum_{j} c_{ij} x_j$. In general, $f_i(\vec{x})$ may be a complicated function due to nonlinear functional responses or other phenomena. Regardless of the exact form of $f_i(\vec{x})$, the ecological fitness can always be linearized near an equilbrium poin, $\vec{x^*}$, in which case the ecological dynamics are approximately described by LV equations. 

MacArthur and Levins \cite{macarthur_limiting_1967} considered an equilibrium point ($\vec{x^*}$) where $x_i \approx 0$ and the linearized ecological fitness of species $i$ is $f_i(\vec{x^*}) \approx K_i -\sum_{j} c_{ij} x_j^*$. If a small number of species $i$ attempt to invade the community, they will be successful if $ K_i -\sum_{j} c_{ij} x_j^* > 0$ but will be unsuccessful if $ K_i -\sum_{j} c_{ij} x_j^* < 0$. If the community has only two species, then $x_j^* = K_j$ and the relationships are $ K_i - K_j c_{ij} > 0$ for successful invasion and $ K_i - K_j c_{ij} < 0$ for unsuccessful invasion. 

The qualitative results should be similar in the case of a stochastic LV dynamics. In the absence of species $i$ the species abundances fluctuate about their mean values ($\vec{x^*}$). If a small number of species $i$ attempt to invade the community, they are most likely to be successful if $ K_i -\sum_{j} c_{ij} x_j^* > 0$, i.e.\ the mean extinction time will be long, but will be unlikely to be successful if $ K_i -\sum_{j} c_{ij} x_j^* < 0$, i.e.\ the mean extinction time will be short.

Now, let's take a look at our proposed immigration and extinction rates. That the rate at which an absent species with $s_i = 0$ becomes a present species with $s_i = 1$ is simply the rate of immigration $\tilde{\lambda}_i$ is straight-forward. In the case of extinction, the rate of extinction is controlled by $ \tilde{K}_i - \sum_j \tilde{K}_j \tilde{c}_{ij} s_j $, for fixed $\w$. This can be compared to $K_i -\sum_{j} c_{ij} x_j^*$, which controls the rate of extinction in the MacArthur-Levins model. If we make a conjecture that, all other things being equal, $x_j^*$ scales proportionally with $K_j$ then we can write $x_j^* = K_j \g_j c_{ij} s_j$ where $\g_j$ is simply the constant of proportionality describing the scaling of $x_j^*$ with $K_j$. Finally, we simply define $\tilde{c}_{ij} = \g_j c_{ij}$ to arrive at our proposed extinction rate. The behavior is such that if $ \tilde{K}_i - \sum_j \tilde{K}_j \tilde{c}_{ij} s_j \gg 0 $ then the rate at which species $i$ goes extinct in the local community will be very slow, whereas if $ \tilde{K}_i - \sum_j \tilde{K}_j \tilde{c}_{ij} s_j \ll 0 $ it will be fast. This behavior is qualitatively similar to MacArthur and Levins analysis of species invasion in LV communities \cite{macarthur_limiting_1967}. 

\begin{figure}[h]
\includegraphics[width=5in]{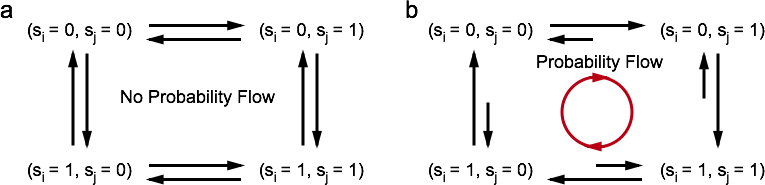}
\caption{ Equilbrium and detailed balance. a) In a system that satisfies detailed balance, the flow of probability into and out of all of the states are equal so that the system remains in equilbrium. b) In a system without detailed balance, there is a non-zero flow between the different configurations of the community, even when the probability distribution reaches a steady-state. 
\label{fig:detailed_balance}}
\end{figure}

\subsection{Equilibrium Distribution}

In general, there is no way of solving Eq.\ \ref{eq:master} to find $P_t(\vec{s})$. However, it is possible to solve for the steady-state equilibrium distribution $P_{PA}(\vec{s})$, i.e.\ the distribution such that $d P_t / dt = 0$, if we impose the following symmetry conditions: $\tilde{\lambda}_i = \tilde{\lambda}$, $\tilde{K}_i = \tilde{K}$, and $\tilde{c}_{ij} = \tilde{c}_{ji}$. The symmetry conditions ensure that transition rates satisfy the principle of detailed balance (see Fig.\ \ref{fig:detailed_balance}). 

Mathematically, it is clear from Eq.\ \ref{eq:master} that if we can find a distribution $P_{PA}(\vec{s})$ such that $ R_i^E(\vec{s} +  \vec{e_i} ) P_{PA}(\vec{s} +  \vec{e_i} )  =  R_i^I(\vec{s}) P_{PA}(\vec{s})$ if $s_i = 0$ and $ R_i^I(\vec{s} -  \vec{e_i} ) P_{PA}(\vec{s} -  \vec{e_i} )  =  R_i^E(\vec{s}) P_{PA}(\vec{s})$ if $s_i = 1$ then $d P_t /dt$ evaulated at $P_{PA}(\vec{s})$ equals zero. Thus, $P_{PA}(\vec{s})$ is a fixed point of the master equation. Moreover, it is possible to show that this fixed point is unique.

This equilbrium distribution is given by:
\be
\label{eq:equil}
P_{PA}(\vec{s}) = \frac{\exp( \sum_i ( \tilde{K} / \tilde{\w} + \ln \tilde{\lambda}) s_i - \frac{\tilde{K}}{2\tilde{\w}} \sum_{(i,j)} \tilde{c}_{ij} s_i s_j)}{Z(\tilde{\l}, \tilde{K}, \tilde{C}, \tilde{\w})} 
\ee
Here, $Z(\tilde{\l}, \tilde{K}, \tilde{C}, \tilde{\w})$ is a normalizing constant such that the total probability sums to one.  It is straight-forward to verify the detailed balance relations, e.g.\ $ R_i^E(\vec{s} +  \vec{e_i} ) P_{PA}(\vec{s} +  \vec{e_i} )  =  R_i^I(\vec{s}) P_{PA}(\vec{s})$, by plugging in equations \ref{eq:master} and \ref{eq:equil}. For example,
\begin{align}
& R_i^E(\vec{s} +  \vec{e_i} ) P_{PA}(\vec{s} +  \vec{e_i} )  =  R_i^I(\vec{s}) P_{PA}(\vec{s}) \nonumber \\
&\Rightarrow \nonumber \\
& \frac{\exp( -( \tilde{K} - \sum_j \tilde{K} \tilde{c}_{ij} s_j )/\w)  \exp( \sum_i ( \tilde{K} / \tilde{\w} + \ln \tilde{\lambda}) s_i - \frac{\tilde{K}}{2\tilde{\w}} \sum_{(i,j)} \tilde{c}_{ij} s_i s_j)}{Z(\tilde{\l}, \tilde{K}, \tilde{C}, \tilde{\w})} = \nonumber \\
& \frac{\tilde{\lambda} \exp( \sum_{j \neq i} ( \tilde{K} / \tilde{\w} + \ln \tilde{\lambda}) s_j - \frac{\tilde{K}}{2\tilde{\w}} \sum_{(k \neq i,j)} \tilde{c}_{kj} s_k s_j)}{Z(\tilde{\l}, \tilde{K}, \tilde{C}, \tilde{\w})} \nonumber \\
&\Rightarrow \nonumber \\
& \exp( -( \tilde{K} - \sum_j \tilde{K} \tilde{c}_{ij} s_j )/\w + \sum_i ( \tilde{K} / \tilde{\w} + \ln \tilde{\lambda}) s_i - \frac{\tilde{K}}{2\tilde{\w}} \sum_{(i,j)} \tilde{c}_{ij} s_i s_j)= \nonumber \\
& \exp( \ln \tilde{\lambda} + \sum_{j \neq i} ( \tilde{K} / \tilde{\w} + \ln \tilde{\lambda}) s_j - \frac{\tilde{K}}{2\tilde{\w}} \sum_{(k \neq i,j)} \tilde{c}_{kj} s_k s_j) \nonumber \\
&\Rightarrow \nonumber \\
& \exp( \ln \tilde{\lambda} + \sum_{j \neq i} ( \tilde{K} / \tilde{\w} + \ln \tilde{\lambda}) s_j - \frac{\tilde{K}}{2\tilde{\w}} \sum_{(k \neq i,j)} \tilde{c}_{kj} s_k s_j)= \nonumber \\
& \exp( \ln \tilde{\lambda} + \sum_{j \neq i} ( \tilde{K} / \tilde{\w} + \ln \tilde{\lambda}) s_j - \frac{\tilde{K}}{2\tilde{\w}} \sum_{(k \neq i,j)} \tilde{c}_{kj} s_k s_j) \nonumber \\
& \blacksquare \nonumber
\end{align}

\subsection{Generality of the Presence/Absence Model}

Our analysis of our presence/absence model in this work depends only on the properties of the steady-state distribution, $\tilde{P}_{PA}(\vec{s})$. Under our symmetry assumptions $\tilde{P}_{PA}(\vec{s}) = P_{PA}(\vec{s})$, although this is not the case in general. Nevertheless, we can generally expand the logarithm of the steady-state distribution in terms of its moments as:
\be
\ln \tilde{P}_{PA}(\vec{s}) = \ln Z + \sum_i h_i s_i + \sum_{i,j} J_{ij} s_i s_j + \sum_{i,j,k} J_{ijk} s_i s_j s_k + \cdots \nonumber
\ee
where the $h$'s and $J$'s are appropriately chosen coefficients. Thus, one way to view our analysis is that we have taken such a series and truncated it after the pairwise term; though there are some subtleties that arise in using statistical mechanics to analyse a steady-state distribution rather than an equilibrium distribution. Moreover, the form of the equilibrium distribution that we analyze (Eq.\ \ref{eq:equil}) can be viewed as a statistical Maximum Entropy model, also known as Binary Markov Random Fields in the statistics and machine learning literature \cite{jaynes1957information,mora2011biological,schneidman2006weak,kindermann1980markov}. These statistical modeling approaches have been used to model phenomena across many different fields. The appearance of phase transitions in these types of models is quite general.

\subsection{Quenched Disorder and the Average Free Energy}

For a particular choice of $\tilde{\lambda}$, $\tilde{K}$, $\tilde{\w}$, and $\tilde{c}_{ij}$, the distribution of the presence/absence of the species in a community is described by the equilibrium probability distribution:
\be
P_{PA}(\vec{s}) = \frac{\exp( \sum_i ( \tilde{K} / \tilde{\w} + \ln \tilde{\lambda}) s_i - \frac{\tilde{K}}{2\tilde{\w}} \sum_{(i,j)} \tilde{c}_{ij} s_i s_j)}{Z(\tilde{\l}, \tilde{K}, \tilde{C}, \tilde{\w})} \nonumber
\ee
where $Z(\tilde{\l}, \tilde{K}, \tilde{C}, \tilde{\w})$ is a normalizing constant such that the total probability sums to one. However, in this work we are not interested in understanding the behavior of the system for a particular choice of parameters, but we instead want to understand the average behavior of the system over many random realizations of the parameters. We will approach this problem by studying the average behavior of $ \ln Z(\tilde{\l}, \tilde{K}, \tilde{C}, \tilde{\w})$ in the limit that the number of species $S$ goes to infinity. 

In physics, the logarithm of the normalization constant of an equilibrium distribution, e.g.\ $ \ln Z(\tilde{\l}, \tilde{K}, \tilde{C}, \tilde{\w})$, is called the free energy of the system. The free energy is an important, and useful, quantity because its derivatives provide the moments of the distribution (i.e.\ average, variance, etc.). Phase transitions correspond to points where one of the derivatives of the free energy (often with respect to stochasticity) is undefined \cite{kadanoff_statistical_2000}. 

There are two ways that one could compute the average free energy: the ``annealed'' average equal to $ \ln \< Z \>$, or the ``quenched'' average given by $ \< \ln Z \>$. Jensen's inequality immediately implies that $\ln \< Z \> \geq \< \ln Z \>$. Analyzing the quenched average of the free energy corresponds to analyzing the properties of typical communities for which the species interactions are drawn from a given probability distribution. Thus, our goal is to calculate the quenched average of $ \ln Z(\tilde{\l}, \tilde{K}, \tilde{C}, \tilde{\w})$. 

In principle, the quenched average free energy can be calculated using ``the replica trick'' \cite{kirkpatrick1978infinite}:
\be
\lim_{S \to \infty} \frac{1}{S} \< \ln Z\> = \lim_{S \to \infty} \frac{1}{S}  \lim_{n \to \infty} \frac{ \< Z^n\> -1}{n} \nonumber
\ee
However, it is sufficient for our purposes to calculate the quenched average free energy using a Random Energy Model approximation \cite{derrida_random-energy_1980,derrida_random-energy_1981,bryngelson1987spin}. 

\subsection{Deriving the Phase Diagram with the Random Energy Model}

Just to be precise, the goal of our Random Energy Model calculation is to compute the quenched average free energy:
\be
 \lim_{S \to \infty} \frac{1}{S} \< F \> = \lim_{S \to \infty} \frac{1}{S} \< \ln Z\> \nonumber
\ee
where the angular brackets denote an average over random realizations of the species interactions. The definition of free energy in themodynamics is $F = U - T H$, where $U$ is the ``internal energy'' (described below), $H$ is the ``entropy'', and $T$ is the ``temperature''.  For a discrete probability distribution, the entropy can be calculated using the Shannon-Gibbs formula $H = \sum_i p_i \ln p_i$ where $p_i$ is the probability of the $i^{th}$ state. That is, the entropy is a measure of fluctuations. We will show that there is a freezing transition in the Random Energy Model where:
\be
 \lim_{S \to \infty} \frac{1}{S} \< H \> = 0 \nonumber
\ee
The calculation follows that in refs.\ \cite{derrida_random-energy_1980,derrida_random-energy_1981,bryngelson1987spin}.

\subsubsection{The Random Energy Approximation}

First, we define a function called the internal energy (or just the energy) as:
\be
\label{eq:energy}
U(\vec{s}) = -\sum_i (1 + \tilde{\Lambda}) s_i + \HF \sum_{(i,j)} \tilde{c}_{ij} s_i s_j
\ee
where $\tilde{\Lambda} = \tilde{\w} \ln \tilde{\lambda} / \tilde{K}$ will be called the immigration potential. This allows us to rewrite the equilibrium distribution as $P_{PA}(\vec{s}) = Z^{-1} \exp( - \tilde{K} U(\vec{s}) / \tilde{\w} )$. So far, we have done nothing except rewrite the equilibrium distribution, i.e.\ Eq.\ \ref{eq:equil}, in a different form that is the typical convention in physics with the identification of temperature $T = \tilde{\w} / \tilde{K}$.

Each of the interaction coefficients, $\tilde{c}_{ij}$, will be treated as an independent random variable with mean $\< \tilde{c}_{ij} \> = \tilde{\mu} / S$ and finite variance $\< \tilde{c}_{ij}^2 \> - \< \tilde{c}_{ij} \>^2 = \tilde{\s}^2 / S$. For example, the interaction coefficients could be randomly drawn from a Gamma distribution as in the LV simulations presented in the Main Text. Note, however, that the precise form of the distribution does not matter for the Random Energy Model calculation. Each of the interaction coefficients in the energy is a random variable and, as a result, the energy itself is also a random variable. In fact, because the energy is a sum of many independent and identically distributed random variables with finite variances we know that it must be approximately normally distributed according to the central limit theorem.  Let $M = \sum_i s_i$ be the number of species present in the community. Since we are interested in the limit where $S$ is large, we will keep only terms $O(S)$. The moments of equation \ref{eq:energy} are
\be
\<U\> \simeq -(\tilde{\Lambda} + 1) M + \HF \tilde{\mu} \frac{M^2}{S}
\ee
and
\be
\<U^2\>-\<U\>^2\simeq \frac{1}{4} \tilde{\s}^2 \frac{M^2}{S}
\ee

It is helpful to simplify our notation by introducing related quantities that do not scale with the number of species, $S$. These ``intensive'' quantities are:
\bea
\label{eq:intensive}
m &=& \frac{M}{S}\\
u(m) & = & \frac{\<U\>}{S} = -(\tilde{\Lambda} + 1) m + \HF \tilde{\mu} m^2  \\
u'(m) & = & \frac{d u}{dm} = -(\tilde{\Lambda} + 1) + \tilde{\mu} m \nonumber \\
v(m) & = & \frac{\<U^2\>-\<U\>}{S} = \frac{1}{4} \tilde{\s}^2 m^2 \nonumber \\
v'(m) & = & \frac{d v}{d m} = \HF \tilde{\s}^2 m \nonumber
\eea
Here, $m = M / S$ is the ``species saturation''; i.e.\ the fraction of species in the regional species pool that are present in the community. 

The energies are approximately normally distributed, so they have a probability density function given by
\be
\label{eq:pdf-energies}
f( U | m ) = \frac{1}{ \sqrt{2 \pi S v(m) }} \exp \left( - \frac{ (U - S u(m) )^2} {2 S v(m) }\right)
\ee
The approximation in using the Random Energy Model to derive the phase diagram for our presense/absence model comes from an assumption that all of the energies are independently distributed. In reality, the correlation between the energies of two configurations of the community, $\vec{s}^{(1)}$ and $\vec{s}^{(2)}$, is a function of their ``overlap'', $\vec{s}^{(1)} \cdot \vec{s}^{(2)}$. These correlations decay as the order of the species interactions is increased; e.g.\ the correlation is smaller if we include interactions between three species than if we only consider pairwise interactions. Thus, our approximation is to neglect these correlations even though we have only included pairwise species interactions. This approximation has been used extensively to model protein folding.

\subsubsection{The Entropy}

The next step in the calculation is the compute the average number of energy levels, $\<n(U)\>$ in the region $[U, U + dU]$. This is given by:
\be
\label{eq:density}
\langle n(U) \rangle  = \sum_{M = 0}^{M = S} \binom{S}{M} f(U| \frac{M}{S} ) 
\ee
Now, we take the limit of a large number of species. When $S$ is large, the binomial cofficient becomes peaked around a maximum value and the sum is dominated by single term. We take the logarithm of Eq. \ref{eq:density} using Stirling's approximation (e.g.\ $\ln S! \approx S \ln S$), drop all terms $O( \ln S )$, and pull the $S$ out front:
\be
\label{eq:log-density-states}
\ln \langle n(U) \rangle = S  \underset{0 < m < 1}{\operatorname{max}} \left\{ -m \ln m - (1-m) \ln (1-m) - \frac{ (u - u(m))^2}{2 v(m)} \right\}
\ee
where $u = U/S$. The maximization over $m$ results from the statement that a single term dominates the sum in Eq.\ \ref{eq:density}.

The Boltzmann formula for the microcanonical entropy, $H(U)$ tells us that $H(U) = \ln \< n(U) \>$. Combining this with the thermodynamic relation:
\be
\frac{\del H} {\del U} = \frac{1}{T} = \frac{ \tilde{K} }{ \tilde{\w}} 
\label{eq:TempDef1}
\ee
allows one to determine the free energy. Specifically,  we have:
\be
\frac{ \del H} {\del U} = - \frac{ (u - u(m))}{v(m)} =  \frac{1}{T}
\ee
Solving for $u$ gives $u = u(m) - v(m) / T$. Plugging this result in Eq.\ \ref{eq:log-density-states} we obtain the entropy:
\be
\label{eq:entropy}
H(m) /S = -m \ln m - (1-m) \ln (1-m) - \HF \frac{v(m) }{T^2}
\ee
and the free energy:
\be
\label{eq:free-energy}
F(m) /S = u(m) - \HF \frac{v(m)}{T} +\frac{1}{T}  (m \ln m + (1-m) \ln (1-m))
\ee
The typical species saturation $m$ is found by minimizing the free energy.

\subsubsection{The Freezing Transition}

The entropy in the Random Energy Model (Eq.\ \ref{eq:entropy}) is a sum of two terms: $-m \ln m - (1-m) \ln (1-m) \geq 0$ and $-\HF v(m) \leq 0$. Thus, there is a possibility that $H(m)/S \leq 0$ if the variance in the energies (that is the variance in the species interactions) is large enough; $v(m) \geq 2(-m \ln m - (1-m) \ln(1-m))$. However, the thermodynamic entropy is a non-negative quantity, so the line $H(m)/S = 0$ defines a ``freezing'' phase transition. The phase boundary can be parameterized as a critical temperature:
\be
\label{eq:critical-temperature}
T_{c} = \sqrt{\frac{v(m)}{2(-m \ln m -(1-m) \ln(1-m) )} }
\ee
Above this critical temperature (which is stochasticity / carrying capacity) the presence/absence of the various species in the community fluctate randomly. Below this critical temperature the presence/absence of the various species in the community freezes into a particular configuraiton determined by the species traits. Now, we will deive the scaling relation presented in the main text from Eq. \ \ref{eq:critical-temperature}. 

\subsection{Scaling Relation for the Phase Boundary}

A simple scaling argument captures the main features of the freezing phase boundary. Before proceeding, it is useful to find a simple scaling for the magnetization (species saturation) $m$. Differentiating Eq.\ \ref{eq:free-energy} with respect to $m$ and setting the equation equal to zero yields:
\be
\label{eq:mag}
m = \HF \left( 1 + \tanh\left( - \HF \frac{1}{T} u'(m) + \frac{1}{4} \frac{1}{T^2} v'(m) \right) \right) \nonumber
\ee
For large $\tilde{\mu}$, we find that $m \sim \tilde{\Lambda} / \tilde{\mu}$. We can combine this relation to derive a scaling relation for the niche-to-neutral phase boundary.

\begin{figure*}[t]
\includegraphics[width=6.5in]{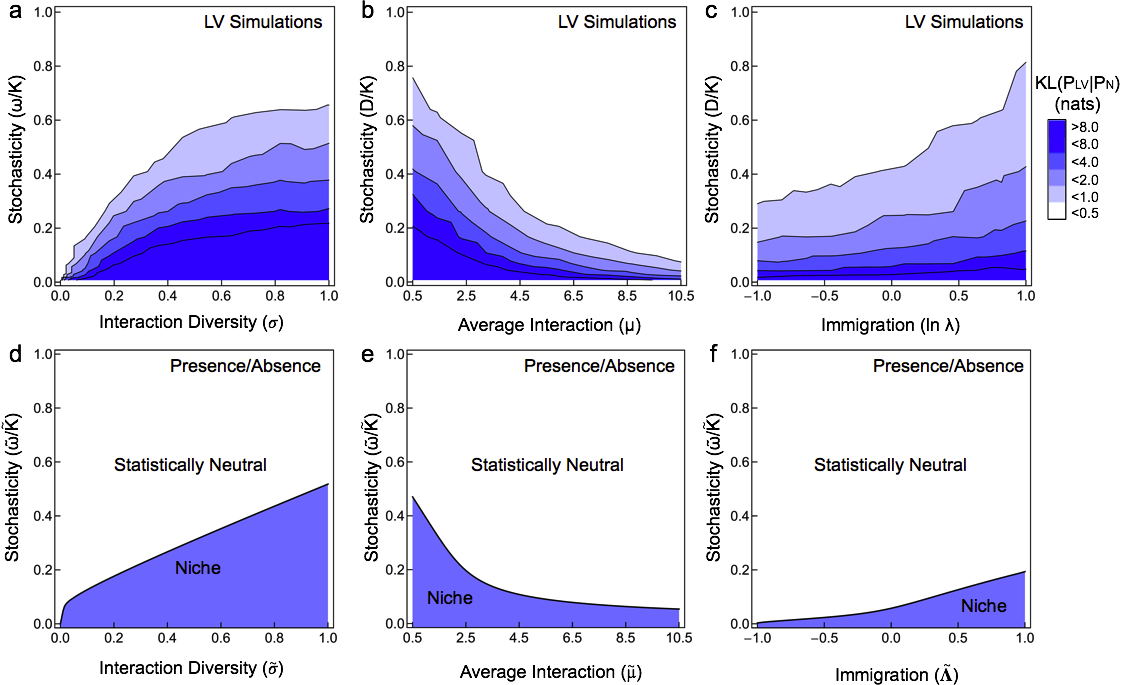}
\caption{ Slices through the niche-neutral phase diagram demonstrate the scaling relation $\w \sim K \ln \lambda \s / \mu$. a-c) The KL-divergence measured during LV simulations while varying interaction diversity ($\s$), the mean interaction strength ($\tilde{\mu}$), and the immigration rate ($\l$). d-f) Phase diagrams calculated from the presence/absence model while varying interaction diversity ($\tilde{\s}$), the mean interaction strength ($\tilde{\mu}$), and the immigration potential ($\tilde{\Lambda}$).  
\label{fig:slices}
}
\end{figure*}

To do so, notice from Eq.\ \ref{eq:TempDef1} that we can define a critical stochasticity
\be
T_{c} = \frac{\tilde{\w_c} }{\tilde{K}} 
\ee
On the other hand,  since $v(m) \sim \sigma^2 m^2$ and $-m\log{m} -(1-m)\log{(1-m)}$ is order 1, Eq.\ \ref{eq:critical-temperature} yields the scaling relation \be
T_c \sim \tilde{\sigma} m \nonumber
\ee
Setting the two equations above equal to each other yields:
\be
\frac{\tilde{\w}}{\tilde{K}} \sim \frac{ \tilde{\s} \times \tilde{\Lambda}} { \tilde{ \mu}} \nonumber
\ee
or, in words:
\be
\frac{ \text{stochasticity} }{ \text{carrying capacity} } \sim \frac{ \text{ immigration} \times \text{interaction diversity} }{ \text{mean interaction strength} } \nonumber
\ee
This result is illustrated in Figure \ref{fig:slices}. In both numerical simulations and analytic calculations, the niche phase is favored when the carrying capacity and interaction diversity are high. By contrast, the neutral phase is favored at low carrying capacities, high stochasticity, and low interaction diversity. These data illustrate that selection dominates community assembly in the niche phase, whereas ecological drift dominates community assembly in the statistically neutral phase. Immigration shifts the phase boundary between the two phases, with high rates of immigration increasing the size of the niche phase relative to the neutral phase.

\subsection{ ``Freezing'' in Lotka-Volterra Communities}

\begin{figure}[h]
\includegraphics[width=3in]{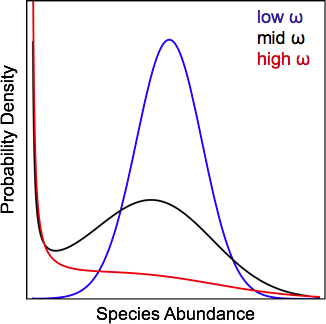}
\caption{Species abundance distributions computed from a single species LV model with different amounts of stochasticty. 
\label{fig:sad}
}
\end{figure}

The Lotka-Volterra model that we study in the main text (and below) is also a stochastic model of population dynamics. It is important to be clear about the effect that the ``freezing'' transition has on these dynamics. The abundances of the species in the community always stochastically fluctuate -- even in the frozen (niche) phase. However, if the fluctuations in the abundance of a species $\<x_i^2\> - \<x_i\>^2$ are much smaller than its mean abundance $\<x_i\>$ then the presence/absence of that species is essentially frozen. These effects are clear even while comparing species abundance distributions across different levels of stochasticity in a single species LV model (see Fig.\ \ref{fig:sad}), though there are important differences between a community with many species and a community with only one species. 

\begin{figure}[h]
\includegraphics[width=5in]{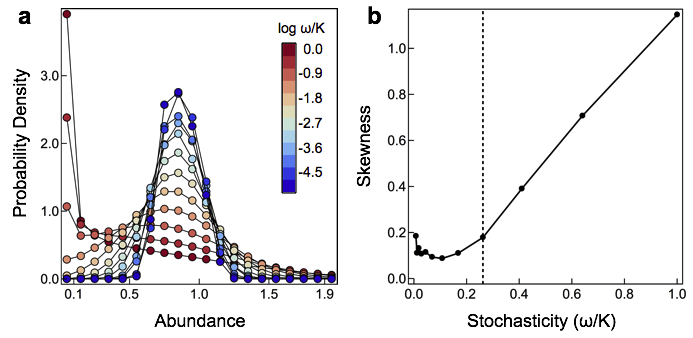}
\caption{Species abundance distributions from LV simulations. a) Species abundance distributions with $\mu = 1.0$, $\s = 0.1$, and $\l = 0.01$. b) Skewness of the species abundance distribution increases beyond the critical stochasticity. 
\label{fig:skewness}
}
\end{figure}

In a diverse community with many species, the niche-to-neutral transition coincides with increase in skewness of the marginal species abundance distribution as a result of the ``unfreezing'' of the species presences/absences (see Fig.\ \ref{fig:skewness}). When the stochasticity increases beyond a critical value, the marginal species abundance distribution transitions from a distribution peaked around $K$ to one peaked at $0$. As a result, the niche-to-neutral transition is accompanied by a loss of biodiversity when the stabilizing selective forces are dominated by stochastic effects. 

\section{Measuring Neutrality from Abundance Distributions}

To test our hypothesis that communities can exhibit a sharp niche-to-neutral phase transition, it is necessary to define an ``order parameter'' that distinguish the niche and neutral phases.  By convention, an order parameter is chosen so that it is zero in one phase, and greater than zero in the other. Recall that the dynamics in the neutral phase are dominated by stochasticity and species abundance distributions in this phase are indistinguishable from those obtained from a neutral model with functionally equivalent species. By contrast, the niche phase is dominated by interactions and species abundance distributions are peaked around the equilibrium value they would have in the absence of stochasticity.

We quantify statistical neutrality in our LV simulations by measuring the distance between the steady-state distributions of species abundances obtained from the LV model ($P_{LV}(\vec{x})$) and purely neutral dynamics  ($P_N(\vec{x})$). The measure of distance that we use in the main text is called the Kullback-Leibler divergence $\text{KL}(P_{LV} || P_{N} ) = \int d \vec{x} P_{LV}(\vec{x}) \ln P_{LV}(\vec{x}) / P_N(\vec{x})$ \cite{kullback_information_1951}. One interpretation of $\text{KL}(P_{LV} || P_{N} )$ is as the amount of information about the true multivariate species abundance distribution (i.e.\ $P_{LV}(\vec{s})$) that is lost by approximating the distribution with one obtained from a neutral model (i.e.\ $P_{N}(\vec{x})$). The KL-divergence ranges from zero to infinity, with $\text{KL}(P_{LV} || P_{N} ) = 0$ implying that the simulated distribution is identical to the distribution obtained under the assumption of neutrality. We study the average of the KL-divergence over many random realizations of the species interactions, i.e.\ $\< \text{KL}(P_{LV} || P_{N} ) \>$. We expect  $\< \text{KL}(P_{LV} || P_{N} ) \>  \approx 0$ in the neutral phase, whereas $\< \text{KL}(P_{LV} || P_{N} ) \> \gg 0$ in the niche phase. 

\begin{figure}[h]
\includegraphics[width=5in]{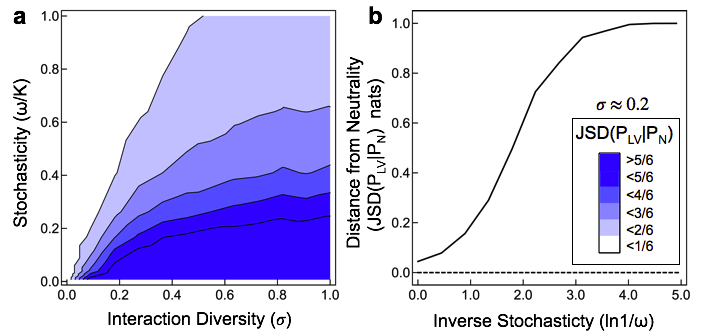}
\caption{Phase diagram and phase transition calculated with the Jensen-Shannon Divergence in LV communities. a) Phase diagram as a function of interaction diversity and stochasticity. b) A slice through the phase diagram demonstrating the sharpness of the transition.  
\label{fig:jsd}
}
\end{figure}

The KL-divergence is not a true distance metric because it is asymmetric and does not satisfy the triangle inequality. There are many alternative measures of disprepancy between probability distributions. One example is the Jensen-Shannon divergence (JSD) given by \cite{lin1991divergence}:
\be
\text{JSD}( P_{LV}, P_{N} ) = \HF( \text{KL}( P_{LV} || \hf ( P_{LV} + P_{N})) + \text{KL}( P_{N} || \hf ( P_{LV} + P_{N})) )\nonumber
\ee
The JSD is symmetric, bounded by $0 \leq \text{JSD}( P_{LV}, P_{N} ) \leq 1$, and although it does not directly satisfy the triangle inequality its square-root ($\sqrt{ \text{JSD}( P_{LV}, P_{N} )}$) does \cite{fuglede2004jensen}. To demonstrate that our results are not very senstive to our choice of metric, Fig.\ \ref{fig:jsd} illustrates the phase diagram and the sharpness of the niche-neutral transition using $\text{JSD}( P_{LV}, P_{N} ) $ instead of $\text{KL}(P_{LV} || P_{N} )$. 

\section{Discussion on Unequal Immigration Rates and Carrying Capacities}

In the main text and in our presence/absence model calculations, we have assumed that all of the species in the community have equal immigration rates and carrying capacities so that species interactions are the only source of variation. The assumption of equal immigration rates is equivalent to assuming that all species are equally abundant in the regional species pool. Of course, this is not generally the case; for example, Hubbell's neutral model results in a regional species pool with abundances that are distributed according to a log-series distribution \cite{hubbell_unified_2001,volkov_neutral_2003}. Additonal simulations discussed below demonstrate that small deviations from the assumption of equal immigration rates and carrying capacities do not affect our main conclusions (see Fig.\ \ref{fig:noisy}). However, it will be important to extend our results to apply in the more general case. In this section, we will provide a brief sketch for how our results can be generalized, but an extensive treatment of the subject will be left for future work. 

The order parameter that we proposed for measuring neutrality based on multivariate species abundance distributions, i.e.\  $\text{KL}(P_{LV} || P_{N} ) = \int d \vec{x} P_{LV}(\vec{x}) \ln P_{LV}(\vec{x}) / P_N(\vec{x})$, is applicable regardless of parameters of the model. In principle, one could explicitly calculate $P_{N}(\vec{x})$ from a specific neutral model \cite{hubbell_unified_2001,volkov_neutral_2003,azaele_dynamical_2006,chisholm_niche_2010}. However, because the parameters of the Lotka-Volterra system and neutral models are different it is not entirely clear to which neutral distribution the comparision should be made. One potential option for quantifying statistical neutrality in communities with unequal immigration rates is to use the the fact that, in a neutral model, the Pearson correlation coefficient between the abundances of species $i$ and $j$ is equal to the correlation coefficient between the abundances of species $k$ and $l$ regardless of the immigration rates. Thus, a distribution with equal correlation coefficients can stand in as a proxie for measuring statistical neutrality. 

Alternatively, we can consider the effects of unequal immigration rates or carrying capacities on the phase diagram of species presence/absence. In general, the stochastic process describing the dynamics of the presence/absence of the species in a community in which the species have unequal immigration rates (or carrying capacities) will not satisfy detailed balance and, thus, never reaches equilibrium. Nevertheless, one can typically assume that there is a steady-state distribution $\tilde{P}_{PA}(\vec{s})$ defined as:
\be
\tilde{P}_{PA}(\vec{s}) = \lim_{t \to \infty} P_{PA}(\vec{s};t) \nonumber
\ee
In general, we can expand the logarithm of the steady-state distribution in terms of its moments as:
\be
\ln \tilde{P}_{PA}(\vec{s}) = \ln Z + \sum_i h_i s_i + \sum_{i,j} J_{ij} s_i s_j + \sum_{i,j,k} J_{ijk} s_i s_j s_k + \cdots \nonumber
\ee
Thus, we can define something like an ``energy'' given by:
\be
U(\vec{s}) =  \sum_i h_i s_i + \sum_{i,j} J_{ij} s_i s_j + \sum_{i,j,k} J_{ijk} s_i s_j s_k + \cdots \nonumber
\ee
and proceed with a calculation analogous to the Random Energy Model. To do so, one only has to relate the means and variances (and potentially covariances) of the $h$'s and $J$'s to the distributions of $\lambda_i$, $K_i$ and $c_{ij}$, and to $\w$. This is a difficult problem to solve exactly, but we can conjecture on the general behaviour. For example, we expect that the mean of $h$ monotonically increases with the mean immigration rate and the mean carrying capacities, but monotonically decreases with increasing $\w$. Likewise, the variance of $h$ should monotonically increase with the variance in the immigration rates and the variance in the carrying capacities, but should decrease monotically with increasing $\w$. Relations of this form provide a general ansatz for studying complex communities, and can be verified numerically. This is an important avenue for future work. 

\section{Additional Lotka-Volterra Simulations}

The qualitative conclusions of our paper hold for a diverse choice of distributions for the LV interaction coefficients. In this section, we present simulations of the LV system (Eq. \ref{eq:LV}) in which the parameters ($K_i$, $\l_i$, and $c_{ij}$) are quenched random variables drawn from various distributions. 

\subsection{Low-rank Interaction Matrix}

\begin{figure}[h]
\includegraphics[width=5in]{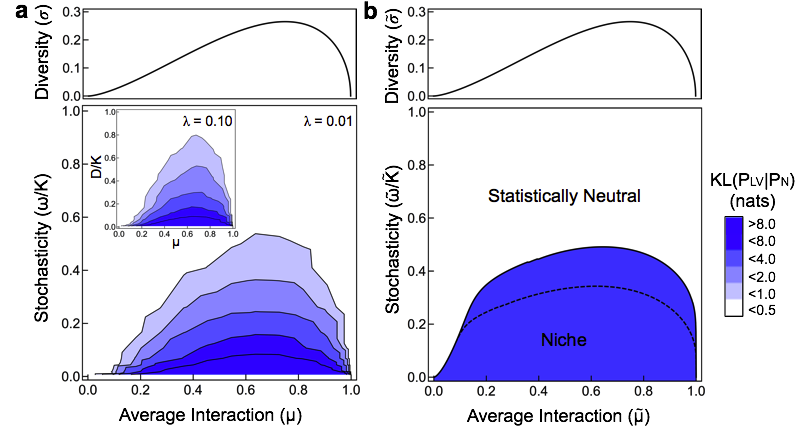}
\caption{ Phase diagram with low-rank interactions. a) Simulations of competitive LV communities with immigration rate $\l = 0.01$ (inset $\l = 0.1$) display two phases: a statistically neutral phase with $\< \text{KL}(P_{LV}||P_{N}) \> \approx 0$ and a niche phase with $\< \text{KL}(P_{LV}||P_{N}) \>) \gg 0$. The critical stochasticity defining the phase boundary is approximately proportional to interaction diversity ($\s$). b) The phase diagram calculated from the PA model with high ($\tilde{\Lambda} = 1.5$; solid) and low ($\tilde{\Lambda} = 0.5$; dashed) immigration potentials has a statistically neutral phase and a niche phase. The niche phase is largest for communities with high interaction diveristy ($\s$) and high rates of immigration.  
\label{fig:low_rank}
}
\end{figure}

Low-rank approximations are often used for inferring matrices because they are less sensitive to noise and overfitting than their full-rank counterparts. Here, we consider a low-rank interaction matrix $c_{ij} = S^{-1} g_i g_j$, where the $g_i$ are independent, Beta distributed random variables on the interval $(0,1)$ with mean $\< g_i \> = \mu$ and variance $\< g_i^2 \> -\< g_i \>^2= \mu(1-\mu)/(1+\nu)$. With this parameterization $\< c_{ij} \> =  S^{-1} \mu^2$ and $\<c_{ij}^2 \> - \< c_{ij} \> ^2 \approx  S^{-2} 2 \mu^3 (1-\mu) / (1+\nu)$. Note that the interaction coefficients ($c_{ij}$) are not independent as in the main text, but are actually correlated. The phase diagrams, as a function of $\mu$, obtained from LV simulations and PA calculations with the low-rank interaction matrix are shown in Fig.\ \ref{fig:low_rank}. 

\begin{figure*}[t]
\includegraphics[width=6in]{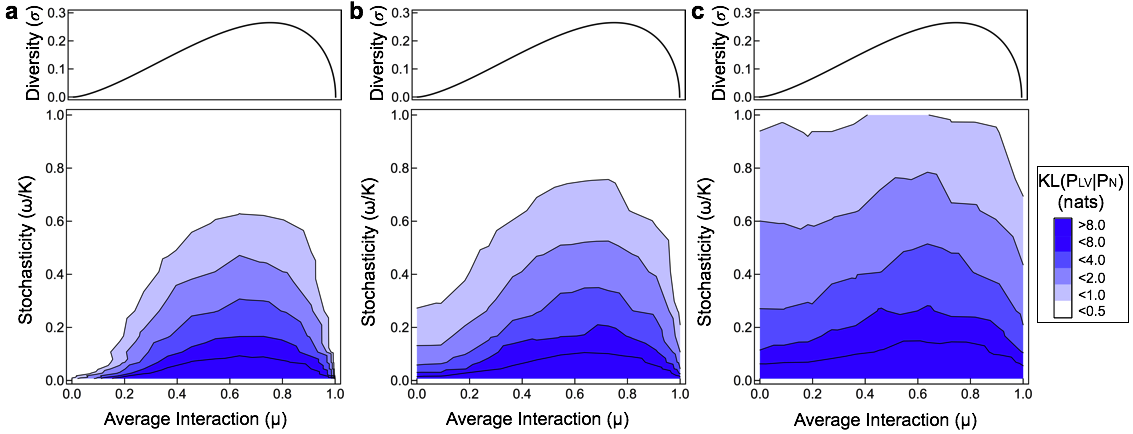}
\caption{Phase diagram of noisy low-rank interactions. a) Simulations of competitive LV communities with 1\% multiplicative noise ($\e = 0.01$). b) Simulations of competitive LV communities with 5\% multiplicative noise ($\e = 0.05$). c) Simulations of competitive LV communities with 10\% multiplicative noise ($\e = 0.10$). In all cases, the immigration rate was $\l = 0.1$.
\label{fig:noisy}
}
\end{figure*}

A similar phase diagram is obtained even if we perturb $\l_i$, $K_i$, and $c_{ij}$ with multiplicative noise drawn from $\mathcal{N}(1,\e)$. (Fig.\ \ref{fig:noisy}). In this case, the interaction matrix is not symmetric, and not necessarily strictly competitive. Moreover, the immigration rates and carrying capacities are no longer equal, that is $\l_i \neq \l_j$ and $K_i \neq K_j$.

\subsection{Exponentially Distributed Interaction Coefficients}

\begin{figure}[h]
\includegraphics[width=3in]{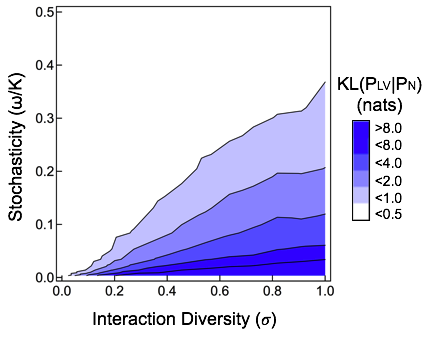}
\caption{Phase diagram of competitive ecosystems with exponentially distributed interaction coefficients. The critical stochasticity scales with functional diversity ($\s$) as predicted from theory, even though the interaction coefficients are drawn from a different distribution. The immigration rate was $\l = 0.1$. 
\label{fig:expo}
}
\end{figure}

In Fig \ref{fig:expo}, we present simulations with constant carrying capacity $K_i = K$, constant immigration rate $\l_i = \l$, and exponentially distributed interaction coefficients $S c_{ij} \sim \text{EXPO}(\s)$. The interaction coefficients have mean $\< S c_{ij} \> = \s$, variance $\< S c_{ij}^2 \> - \< S c_{ij} \>^2 = \s^2$, and probabiility density function
\be
f(S c_{ij}) = \frac{1}{\s} e^{ - S c_{ij} / \s} \nonumber
\ee

\subsection{Pareto Distributed Interaction Coefficients}

\begin{figure}[h]
\includegraphics[width=3in]{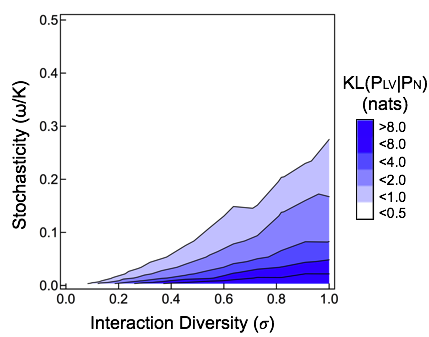}
\caption{Phase diagram of competitive ecosystems with Pareto distributed interaction coefficients. The critical stochasticity scales with functional diversity ($\s$) as predicted from theory, even though the interaction coefficients are drawn from a different distribution. The immigration rate was $\l = 0.1$. 
\label{fig:pareto}
}
\end{figure}

In Fig \ref{fig:pareto}, we present simulations with constant carrying capacity $K_i = K$, constant immigration rate $\l_i = \l$, and Pareto distributed interaction coefficients $S c_{ij} \sim \text{PARETO}(\s) $. The interaction coefficients have mean $\< S c_{ij} \> \approx \frac{\s}{1+\s -\s^2/2} $, variance $\< S c_{ij}^2 \> - \< S c_{ij} \>^2 \approx \s^2$, and probabiility density function
\be
f(S c_{ij}) = \frac{2+ \s^{-1}-\s/2}{ (1+ S c_{ij})^{3 + \s^{-1}-\s/2}} \nonumber
\ee
We require $\s < \sqrt{2}$ in order for this distribution to have a finite second moment.

\section{Island Biogeography}

\begin{figure*}[t]
\includegraphics[width=6.5in]{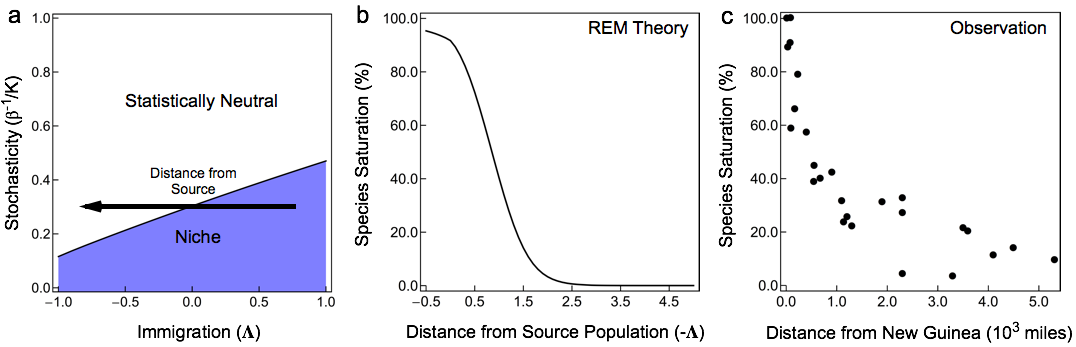}
\caption{ Spatial variation in immigration and island biogeography. a) Islands at different distances from the mainland are located in different regions of the phase diagram because the immigration potential ($\tilde{\Lambda}$) decreases with distance from the source population. b) The percentage of species inhabiting an island (its saturation) decreases with distance (-$\tilde{\Lambda}$) from the source population in the PA model. The PA model calculations are for $\tilde{\mu} = 0.5$, $\tilde{\s} = 0.5$, and $\tilde{\w}/\tilde{K} = 0.3$. c) Similarly, species saturation decreases with distance from the source population in islands off the coast of New Guinea \cite{macarthur1963equilibrium}.  
\label{fig:island}
}
\end{figure*}

The qualitative features of the equilibrium behavior of a community are described by its position on the phase diagram just as the state of water can be characterized as a solid, liquid, or gas depending on pressure and temperature. The ecological consequences of the niche-to-neutral transition can be illustrated by comparing communities that correspond to different regions of the phase diagram, or by the studying how the characteristics of a community change in response to a disturbance that moves the community from one phase to another. 

According to the theory of island biogeography, the rate of immigration to an island decreases with its distance from the source population such that $\Lambda \sim - \text{distance}$ \cite{macarthur1963equilibrium}. Therefore, a chain of islands at different distances from the mainland traces a path along the immigration axis of the phase diagram (Fig.\ \ref{fig:island}a). The percentage of species from the source population that inhabit an island, its ``saturation'' $100 \times \sum_i s_i / S$, decreases with the distance of the island from the mainland. As shown in Figures \ref{fig:island}b-c, our model can reproduce the qualitative trend in saturation as a function of distance famously observed for islands off the coast of New Guinea \cite{macarthur1963equilibrium}. In addition, our model predicts that there is a distance that defines a sharpe niche-to-neutral transition.

\section{Methods Used in Numerical Simulations}

We simulated Hubbell's neutral model with a local community of $J$ individuals connected to an infinitely large metacommunity containing $S = 50$ equally abundant species \cite{hubbell_unified_2001}. In each timestep, with probability $\l$, an individual randomly drawn from the metacommunity replaced a randomly chosen individual in the local community, or with probability $1-\l$, one randomly chosen individual in the local community replaced another randomly chosen individual in the local community. The simulations were run for $5  \times 10^{7}$ steps. Ten simulations were run for each set of parameters, and the results were averaged.

We studied well-mixed, competitive communities described by a system of stochastic Lotka-Volterra (LV) equations given by
\be
\label{eq:LV}
\frac{dx_i (t)}{dt} = \lambda_i + x_i (K_i - x_i - \sum_{j \neq i} c_{ij} x_j) + \sqrt{\w x_i} \eta_i (t) 
\ee
LV simulations with $S=50$ species were performed over the parameter ranges specified in the figure legends. The system of equations (Eq. \ref{eq:LV}) was forward integrated using the Milstein method, as
\be
\begin{split}
x_i(t+\d t) &= x_i(t) + \lambda \d t + x_i(t) \left(K_i -x_i(t) -\sum_{j\neq i} c_{ij} x_j(t) \right) \d t \\
&  \quad + \sqrt{ \w x_i(t)} \d W + \frac{\w}{4} \left( (\d W)^2 -\d t \right) \nonumber
\end{split}
\ee
where $\d W \sim \sqrt{\d t} \mathcal{N}(0,1)$, and $\mathcal{N}(0,1)$ denotes a standard normally distributed random variable. In each case, the competition coefficients were sampled randomly, then the stochastic Lotka-Volterra equations (Eq. \ref{eq:LV}) were forward integrated for $5 \times 10^7$ steps of size $\delta t = 0.005$ using the Milstein method \cite{milstein1995numerical}. Ten simulations were run for each set of parameters, and the results were averaged.

\subsection{Quantifying Statistical Neutrality}
The degree of statistical neutrality, $\text{KL}(P_{LV}||P_{N})$, was measured using Gaussian approximations for $P_{LV}(\vec{x})$ and $P_{N}(\vec{x})$. The mean abundance of species $i$ was calculated from a simulation of length $\tau$ using
$ \<x_i\> = \tau^{-1} \sum_{t=0} ^{t=\tau} x_i(t)$,
and the covariance between species $i$ and $j$ was
$ C_{ij} = \tau^{-1} \sum_{t=0}^{t=\tau} (x_i(t) - \<x_i\>)(x_j(t) - \<x_j\>)$. The corresponding statistically neutral moment estimates were
$\bar{ x }= S^{-1}\sum_{i=1}^{i=S} \<x_i\>$, 
$\bar{C_{ii}} = S^{-1} \sum_{i=1}^{i=S} C_{ii}$, 
and $\bar{C_{ij}} = 2 S^{-1} (S-1)^{-1} \sum_{i < j} C_{ij}$ for the mean, variance and covariance, respectively. Thus, $P_{LV}(\vec{x})$ (or $P_{N}(\vec{x})$) is a multivarate Gaussian distribution with mean $\< \vec{x} \>$ (or $\bar{\vec{x}}$) and covariance matrix $\mat{C}$ (or $\bar{\mat{C}}$). The KL-divergence was calculated as:
\bea
\text{KL}(P_{LV}||P_{N}) = \frac{1}{2} (\text{tr}(\bar{C}^{+} C) - \text{rk}(\bar{C}) + (\bar{x}  \nonumber \\
 - \<x\>)^{'} \bar{C}^{+} (\bar{x} - \<x\>) - \ln |C|_{+} + \ln |\bar{C}|_{+}
 \eea
Here, $|M|_{+}$ denotes the pseudo-determinant of the matrix $M$, $M'$ denotes its transpose, $M^{+}$ denotes its pseudo-inverse, and $\text{rk}(M)$ denotes its rank. The pseudo determinant and inverse were used because, in Hubbell's model, the constraint that the number of individuals in the local community is fixed reduces the degrees of freedom by $1$. To calculate the degree of statistical neutrality for a set of parameters for the competition coefficients, we averaged the KL-divergence over 10 independent draws of the interaction coefficients ($c_{ij}$)

\section{Related Literature}

This section provides some related references on phase transitions in populations. In particular, there are a number of papers on phase transitions in systems of replicator equations (which are closely related to Lotka-Volterra models) with disordered species traits \cite{diederich1989replicators,chawanya2002large,de2003replicators,poderoso2005random,de2001extinctions,galla2006random}. The Random Energy Model has also been applied to study phase transitions in evolution (for a review see \cite{neher2011statistical}). 

\bibliography{refs_7-30-13}

\end{document}